\documentclass[journal=nalefd,manuscript=article]{achemso}
\mciteErrorOnUnknownfalse

\usepackage{chemformula} 
\usepackage[T1]{fontenc} 
\usepackage{siunitx} 
\usepackage[version=4]{mhchem} 

\author{David Lamprecht}
\affiliation[TU VIENNA]
{Institute for Microelectronics, TU Wien, Gußhausstraße 25-29, 1040 Vienna, Austria}
\alsoaffiliation[University of Vienna]
{University of Vienna, Faculty of Physics, Boltzmanngasse 5, 1090 Vienna, Austria}
\email{lamprecht@iue.tuwien.ac.at}

\author{Anna Benzer}
\affiliation[TU VIENNA]
{Institute for Microelectronics, TU Wien, Gußhausstraße 25-29, 1040 Vienna, Austria}
\author{Manuel Längle}
\affiliation[University of Vienna]
{University of Vienna, Faculty of Physics, Boltzmanngasse 5, 1090 Vienna, Austria}
\author{Mate Capin}
\affiliation[TU VIENNA]
{Institute for Microelectronics, TU Wien, Gußhausstraße 25-29, 1040 Vienna, Austria}
\author{Clemens Mangler}
\affiliation[University of Vienna]
{University of Vienna, Faculty of Physics, Boltzmanngasse 5, 1090 Vienna, Austria}
\author{Toma Susi}
\affiliation[University of Vienna]
{University of Vienna, Faculty of Physics, Boltzmanngasse 5, 1090 Vienna, Austria}
\email{toma.susi@univie.ac.at}

\author{Lado Filipovic}
\affiliation[TU VIENNA]
{Institute for Microelectronics, TU Wien, Gußhausstraße 25-29, 1040 Vienna, Austria}
\author{Jani Kotakoski}
\affiliation[University of Vienna]
{University of Vienna, Faculty of Physics, Boltzmanngasse 5, 1090 Vienna, Austria}
\email{jani.kotakoski@univie.ac.at}

\title[An \textsf{achemso} demo]
  {Uncovering the atomic structure of substitutional platinum
dopants in MoS$\mathrm{_2}$ with single-sideband ptychography}

\keywords{4D-STEM, Ptychography, MoS$_2$, Pt Dopants, 2D materials, SSB}

\begin{document}



\begin{abstract}
We substitute individual Pt atoms into monolayer \ce{MoS_2} and study the resulting atomic structures with single-sideband ptychography (SSB) supported by \textit{ab initio} simulations. We demonstrate that while high-angle annular dark-field (HAADF) scanning transmission electron microscopy (STEM) imaging provides excellent \textit{Z}-contrast, distinguishing some defect types such as single and double sulfur vacancies remains challenging due to their low relative contrast difference. However, SSB with its nearly linear \textit{Z}-contrast and high phase sensitivity enables reliable identification of these defect configurations as well as various Pt dopant structures at significantly lower electron doses. Our findings uncover the precise atomic placement and highlight the potential of SSB for detailed structural analysis of dopant-modified 2D materials while minimizing beam-induced damage, offering new pathways for understanding and engineering atomic-scale features in 2D systems.
\end{abstract}

\section{Introduction}
Due to their high surface-to-volume ratio, 2D materials are promising candidates as active material for future catalytic and gas-sensing applications. Particularly \ce{MoS_2}, which as a monolayer is an intrinsic direct band gap semiconductor, has attracted significant interest. However, a major limitation of \ce{MoS_2} as a catalytic material is the relative chemical inertness of its basal plane which severely restricts its potential use-cases~\cite{cao_roadmap_2021}. To overcome this problem, various material modification methods like surface metal decoration~\cite{ge_dual-metallic_2021, shi_selective_2013}, defect-engineering~\cite{wang_single-atom_2020, madaus_highly_2018}, or the assembling of heterostructures with other 2D materials~\cite{woods_one-step_2016, zhao_edge-enriched_2022} have been proposed and experimentally verified. Substitutional doping, where a single heteroatom replaces one or more atoms in the lattice, is considered a modification method of particular interest~\cite{filipovic_application_2022} due to its simplicity and potentially high selectivity. Replacement of S atoms has been reported for over half of the elements on the periodic table~\cite{sovizi_single_2022}, but atomic resolution confirmation of this incorporation remains scarce.

Substitution to chalcogene sites has been achieved either with direct incorporation during chemical vapor deposition (CVD) growth, e.g., O~\cite{tang_situ_2020}, Va~\cite{zhang_vanadium-doped_2021}, alloying with other chalcogenes, e.g., Se~\cite{gong_band_2014}, post-growth plasma implantation, e.g., N~\cite{azcatl_covalent_2016}, Cl,~\cite{kim_effect_2020} or various hydrothermal methods, e.g., Rh~\cite{meng_distance_2020}, W~\cite{rong_restructuring_2022}. Evidence for the substitution of Mo atoms is limited to bottom-up methods and mostly performed by adding metals to the precursor during CVD growth, e.g., Fe~\cite{li_enhanced_2020}, Ta~\cite{li_controllable_2022}, and hydrothermal growth methods, e.g., Pd~\cite{liu_optimization_2024}. Notably, there are few reports of implanting precious metals like Pt into \ce{MoS_2}, despite various theoretical predictions regarding the potential of Pt-doped \ce{MoS_2} for gas-sensing and catalysis~\cite{chen_noble_2018,burman_substitutional_2021, li_synergistic_2021}.

In Ref.~\citenum{li_atomic_2017}, Li et al. studied single Pt atoms on \ce{MoS_2} by separating individual atoms from clusters using the electron beam of an aberration-corrected scanning transmission electron microscope (STEM). They were able to successfully implant single Pt atoms into S vacancy sites and study their dynamics under the electron beam. However, their substitution method is barely scalable and not trivially adaptable to other elements or atomic sites. Several other studies have claimed selective substitution of Mo atoms with Pt~\cite{li_synergetic_2018, burman_substitutional_2021, shi_phase-dependent_2023} atoms, but none have provided atomic-resolution confirmation, which is essential to distinguish between true lattice incorporation and mere surface decoration.

In this study, we extend our previously published two-step implantation method, originally demonstrated for implanting graphene with Au~\cite{trentino_single_2024,trentino_two-step_2022}, Fe, Ag, Ti,~\cite{trentino_single_2024} and Al,~\cite{trentino_single_2024,zagler_beam-driven_2022} to \ce{MoS_2}. First, we introduce defects into monolayer \ce{MoS_2} using low-energy He ion irradiation with a plasma source~\cite{huang_selective_2019}. Subsequently we fill the vacancies with single Pt atoms stemming from an evaporation source. For structural analysis of the modified material, high-angle annular dark-field (HAADF)-STEM and simultaneous 4D-STEM imaging is carried out. The resulting 4D data stacks are used to reconstruct the phase information using the single-sideband ptychography (SSB)  algorithm~\cite{pennycook_efficient_2015}. 

While in HAADF imaging the intensity of an individual atom scales with the atomic number~\cite{krivanek_atom-by-atom_2010} \textit{Z}$^{1.64}$, the phase contrast in SSB is approximately linear to the amplitude of the projected potential~\cite{yucelen_phase_2018} \textit{Z}, which results in an approximate linear dependence on \textit{Z} for single atoms. This allows the simultaneous and precise imaging of neighboring heavy and light atoms with SSB, which is needed for the analysis of our structures~\cite{jiang_electron_2018,wen_simultaneous_2019, loh_electron_2025}. We show that SSB not only allows to reliably differentiate between different vacancy structures but also between Pt atoms trapped in single ($\mathrm{V_{1S}}$) and double S vacancies  ($\mathrm{V_{2S}}$), a distinction that is challenging in HAADF-STEM due to their low contrast difference. 
To obtain further evidence for the correct assessment of the defect structures, we compare the obtained experimental images with image simulations and study the vacancy-mediated substitution process using density functional theory (DFT). 
Overall, as our method relies on the filling of vacancies by adatoms, our results demonstrate a pathway for controlled substitutional doping of MoS$_2$ with arbitrary elements.
\section{Results and discussion}

CVD-grown \ce{MoS_2} samples were transferred from the \ce{SiO_2} substrate to Quantifoil Au TEM grids and subsequently introduced into a interconnected UHV system~\cite{mangler_materials_2022}. The sample contains large atomically clean areas with a low density of intrinsic defects when investigated with STEM. After initial imaging, the samples were transferred in UHV to the sample manipulation chamber and exposed to He ions with a kinetic energy of 171 $\pm$ 21 eV, which is the intrinsic kinetic energy for He ions of the used plasma source at a He partial gas pressure of ca. 2.5 $\times 10^{-5}$ mbar. The sample treatment is illustrated in Fig.~1, including atomic-resolution images at the different stages. To produce a defect in the sample surface, an impinging ion has to be able to transfer enough kinetic energy to the surface atom to overcome its displacement threshold. Assuming a simple fully elastic knock-on event with a maximum energy transfer, a He ion needs a minimum kinetic energy of 130.2 eV to overcome the displacement threshold of ca. 20 eV~\cite{komsa_two-dimensional_2012} of a Mo atom in MoS$_2$. As the displacement threshold for S atoms is only ca. 6.9 eV~\cite{yoshimura_quantum_2023}, a minimum kinetic energy of 17.9 eV is already enough to produce a $\mathrm{V_{1S}}$ defect. Therefore the He ions with a kinetic energy of 171 eV ions are able to introduce defects into both sublattices.

\begin{figure}[htp]
    \centering
    \includegraphics[width= 16 cm]{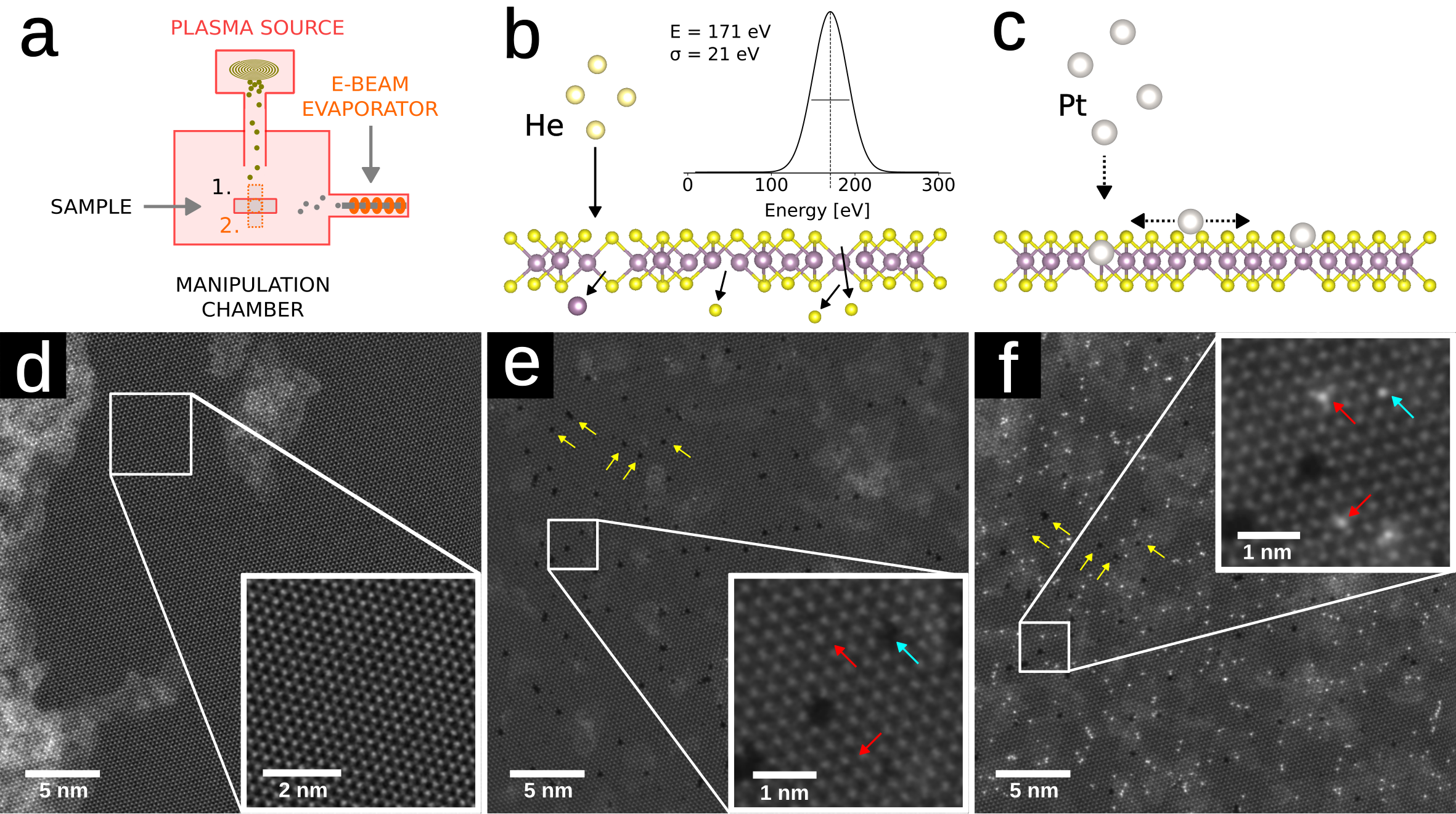}
    \caption{a) Schematic illustration of the sample manipulation chamber used for the study. In step 1 the sample is subjected to the ion beam, in step 2 the sample is in the field of view of the e-beam evaporator. b) Schematic illustration of the defect-engineering process. The inset shows the beam energy profile of the He ions. (The full \textit{dI}/\textit{dV} curve can be found in Supplemental Materials Fig. S1.) c) Schematic illustration of the single atom evaporation process. d) HAADF-STEM image of a clean \ce{MoS_2} area before modification steps (not the same area as in the following images). e) HAADF-STEM image \ce{MoS_2} after 10 min irradiation with He ions. The red and turquoise arrows in the inset indicate $\mathrm{V_{1S}}$ (red) and $\mathrm{V_{Mo}}$ (turquoise) and defect sites that will be filled with Pt atoms, the yellow arrows mark the same defect features before and after Pt evaporation. f) HAADF-STEM image of roughly the same area and field-of-view as in panel e (see yellow arrows); the sites filled with Pt atoms are indicated with the red and turquoise arrows in the inset. }
    \label{Fig 1}
\end{figure}

After 10 min of ion irradiation with an estimated total fluence of 1.25 × 10$^{13}$ cm$^{-2}$ the samples were transported under UHV to the microscope to image the defect structures. After imaging, Pt atoms were evaporated onto the sample using an e-beam evaporator, while keeping the sample under UHV. Nearly exactly the same sample area is shown in Fig.~\ref{Fig 1}e before and in Fig.~\ref{Fig 1}f after the evaporation. It is evident that Pt atoms are incorporated into the \ce{MoS_2} lattice, occupying the former $\mathrm{V_{1S}}$, $\mathrm{V_{2S}}$  and molybdenum vacancy ($\mathrm{V_{Mo}}$) sites.

\begin{figure}[htp]
    \centering
    \includegraphics[width= 16 cm]{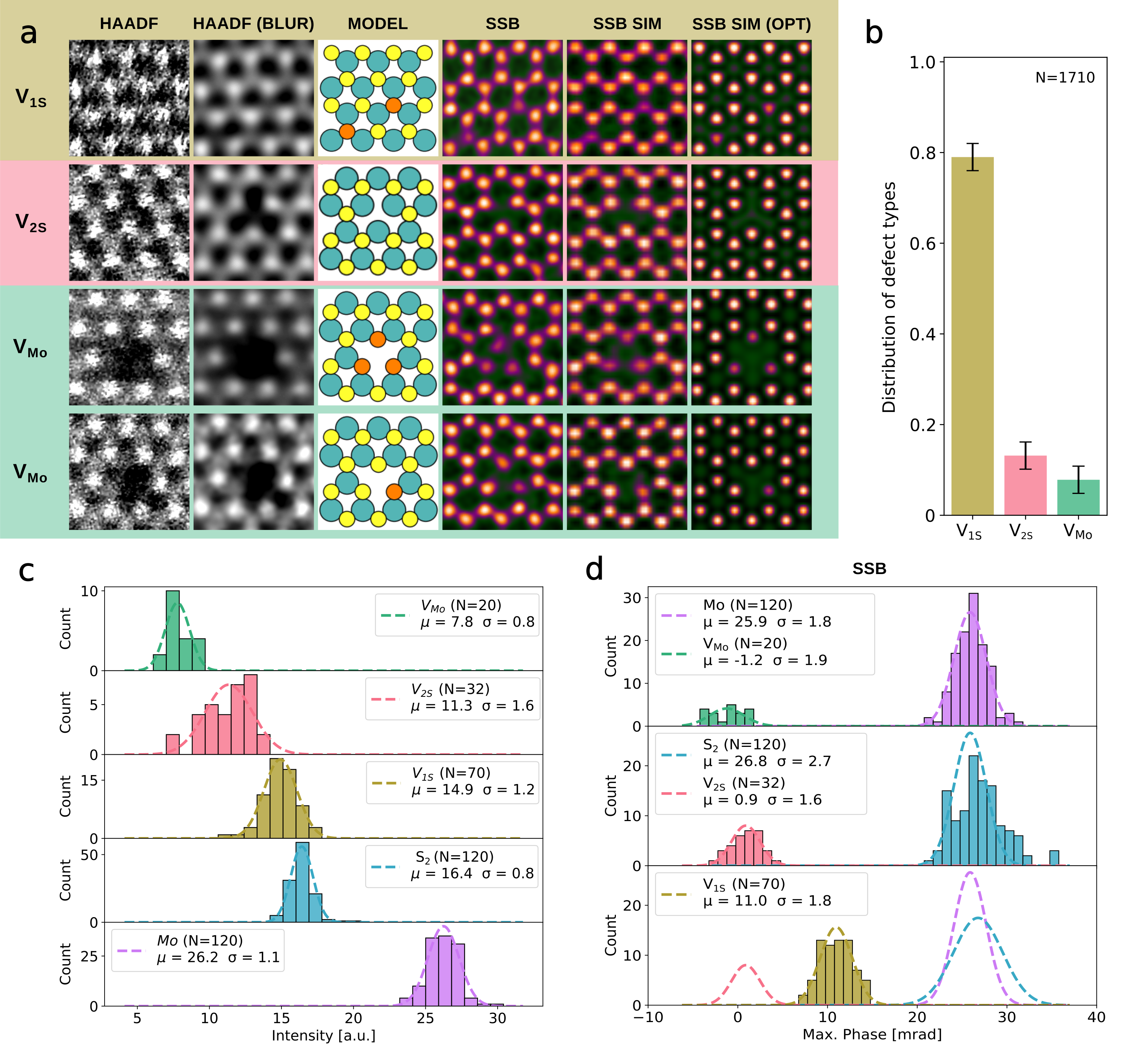}
    \caption{a) HAADF-STEM images (field of view ca. 1 nm) of defect structures without and with Gaussian blurring, atomic models of the imaged structures, SSB reconstructions of the phase information at the same location as well as simulations of the SSB images corresponding to experimental parameters. The last column contains simulations of the SSB images under perfect conditions (unlimited dose, no residual aberrations). b) Relative occurrence of different defect types in the defect-engineered \ce{MoS_2} based on SSB and HAADF images. The uncertainty in the columns is based on the variation between observed images. c) Histograms of HAADF intensities at the Mo and S$\mathrm{_2}$ sublattice sites. The type of the S vacancies (no vacancy, $\mathrm{V_{1S}}$, $\mathrm{V_{2S}}$) are determined using the SSB intensity at the respective S sublattice site. d) Histograms of SSB phase values at the same atomic sites as in c), together with Gaussian fits of the phase distribution of all structures. The $\mu$ in the legends indicates the center of the Gaussian fits, the $\sigma$ is the standard deviation of the respective Gaussian. \textit{N} is the number of cases for each histogram.}
    \label{Fig 2}
\end{figure}

HAADF and SSB images of defect structures are shown in Fig.~\ref{Fig 2}a together with corresponding image simulations conducted with \textit{ab}TEM~\cite{madsen_abtem_2021} based on relaxed atomic models. A comparison of the HAADF intensities of $\mathrm{V_{1S}}$ and $\mathrm{V_{2S}}$ and the pristine S$\mathrm{_2}$ sites shows that it is difficult to differentiate between the pristine S$_2$ sublattice and S vacancies in HAADF imaging, requiring high doses and high magnification, which come with the disadvantage of introducing additional defects during the imaging process. Fortunately 4D-STEM ptychography has a much higher contrast and dose efficiency compared to HAADF images~\cite{jiang_electron_2018}. For 4D data collection, stacks with 512$\times$512 real-space pixels were collected with a dwell time of 20 $\mathrm{\mu s}$ and an average dose of ca. $1 \times 10^{5}$  $\mathrm{e^-}$/\AA$^2$. The phase information was retrieved using the single-sideband (SSB) method with post-acquisition aberration correction as described in Ref.~\citenum{pennycook_efficient_2015}. 

The last two rows of Fig.~\ref{Fig 2}a contain examples of defect clusters around a $\mathrm{V_{Mo}}$ site. While in the HAADF images, the neighborhood of the Mo vacancy is quite ambiguous and the number of neighboring S vacancies is hardly determinable, the SSB images clearly show a well-defined atomic structure. Mo-S vacancy structures with one to five missing sulfur atoms and a number of different vacancy configurations can be observed. Due to the relatively low ion fluence used in the experiments, the appearance of these vacancy structures are most likely due to a single impact and following collision cascades. 

Figure~\ref{Fig 2}b shows the ratio of the defect numbers derived from both large-scale HAADF imaging and small-scale phase reconstructions. Over 80\% of the introduced defects are $\mathrm{V_{1S}}$, whereas $\mathrm{V_{2S}}$ and $\mathrm{V_{Mo}}$ contribute with ca. 12\% and 8\% to the overall defect density of $1.3 \pm 0.4$ defects per nm$\mathrm{^2}$, in agreement to what we would expect at this ion energy~\cite{ghorbani-asl_two-dimensional_2017}.

Figure~\ref{Fig 2}c-d contains histograms of the HAADF intensity and phase maxima at the same atomic positions in the defect-engineered \ce{MoS_2} as well as Gaussian fits of the distributions. The HAADF intensity distributions of pristine S$\mathrm{_2}$, $\mathrm{V_{2S}}$ and $\mathrm{V_{1S}}$ overlap significantly with each other and form a near-uniform distribution. By contrast, SSB images exhibit significantly larger phase ratios between the atomic columns, enabling more precise defect identification. The mean phase of the $\mathrm{V_{1S}}$ site is 11.0 $\pm$ 0.6~mrad, while the mean phase of the S$\mathrm{_2}$ site is 26.8 $\pm$ 0.7~mrad, which leads to an average phase ratio of 2.4 between the two cases, allowing for precise discrimination between defective and pristine sites in the sulfur sublattice. Additionally, the $\mathrm{V_{2S}}$ site with its mean phase of 0.9 $\pm$ 1.3~mrad can be easily discerned from both $\mathrm{V_{1S}}$ and $\mathrm{S_2}$. The uncertainties given for the mean phases correspond to the 95\% confidence interval.

The distributions of $\mathrm{V_{2S}}$ and $\mathrm{V_{Mo}}$ are both centered near a phase of zero. Even though this seems to match optically with the simulated images, there is a significant difference visible in the line profiles of these structures (Supplementary Material Fig.~S3). Unlike in HAADF imaging, the SSB phase displays a negative phase halo around a single atom which converges to the background value of zero phase after some distance~\cite{hofer_reliable_2023}. In the case of \ce{MoS_2}, the negative halos of the six atomic columns around one hexagon nearly overlap, creating a deep phase trench with a small spike in the center of the hexagon. As only few of the observed $\mathrm{V_{2S}}$ have a negative phase maximum we suspect that the sites are actually filled with light elements, which has already been discussed by Yin Wen~\cite{wen_structural_2021}. Supplementary Material Fig.~S3c-d contains image simulations and corresponding line profiles of $\mathrm{V_{2S}}$ doped with C and O, which are in a good agreement with the observed phase maxima at the supposed $\mathrm{V_{2S}}$ sites. Additional evidence stems from the fact that an unfilled $\mathrm{V_{2S}}$ would be subject to a lattice contraction of up to 12\%~\cite{wang_detailed_2016}, which is not observed here. These substitutions are most likely C atoms from the hydrocarbon contamination which diffuse freely on the \ce{MoS_2} surface due to their low diffusion barriers (0.56 eV for C~\cite{park_unveiling_2023}, in comparison 1.92~eV for O,~\cite{farigliano_initial_2020}), before they fall into an energetically more favorable $\mathrm{V_{2S}}$ vacancy site (binding energy of 4.5~eV~\cite{park_unveiling_2023}).

Even though similar reasoning could be applied to the observed phase maxima at $\mathrm{V_{Mo}}$ sites, the line profile analysis in Supplementary Material Fig.~S4a-b shows that the center position of the $\mathrm{V_{Mo}}$ sites have a local maximum which is much lower than expected for $\mathrm{V_{Mo}}$ filled with a C atom in Fig.~S4d. Therefore, we think that the maximum values plotted in Fig. 2c indeed correspond to an unfilled $V_{Mo}$. A notable exception is shown in
Supplementary Material Fig. S4c, where the $V_{Mo}$ is clearly filled with an heteroatom, which we assume to be a S atom based on the very similar contrast to the neighboring S atoms.

Due to the very similar phases of pristine Mo with S$_2$ sites, as well as $V_{Mo}$ with $V_{S2}$ sites, it is difficult to determine the sublattices from SSB images alone. This is however not an issue as the concurrent signal from HAADF (or virtual ADF) imaging provides sufficient contrast to distinguish between the Mo and S$\mathrm{_2}$ sublattices.  While in HAADF images the presence of thin hydrocarbon contamination on the \ce{MoS_2} surface and at vacancy sites is indicated only by a low increase in background intensity, SSB reconstruction allows the visualization of individual carbon atoms in the vacancy sites and, to some degree, also on the \ce{MoS_2} surface (for an example see Supplementary Material Fig.~S5). The phase variations introduced due to these light element adatoms, together with the overlapping negative halo effect on nearby atoms~\cite{hofer_reliable_2023}, residual aberrations, scan distortions, and shot noise due to the limited dose~\cite{jiang_electron_2018}, are probably the main contributors to the observed phase variations of up to 3 mrad in all phase measurements. 

\begin{figure}[htp]
    \centering
    \includegraphics[width= 14 cm]{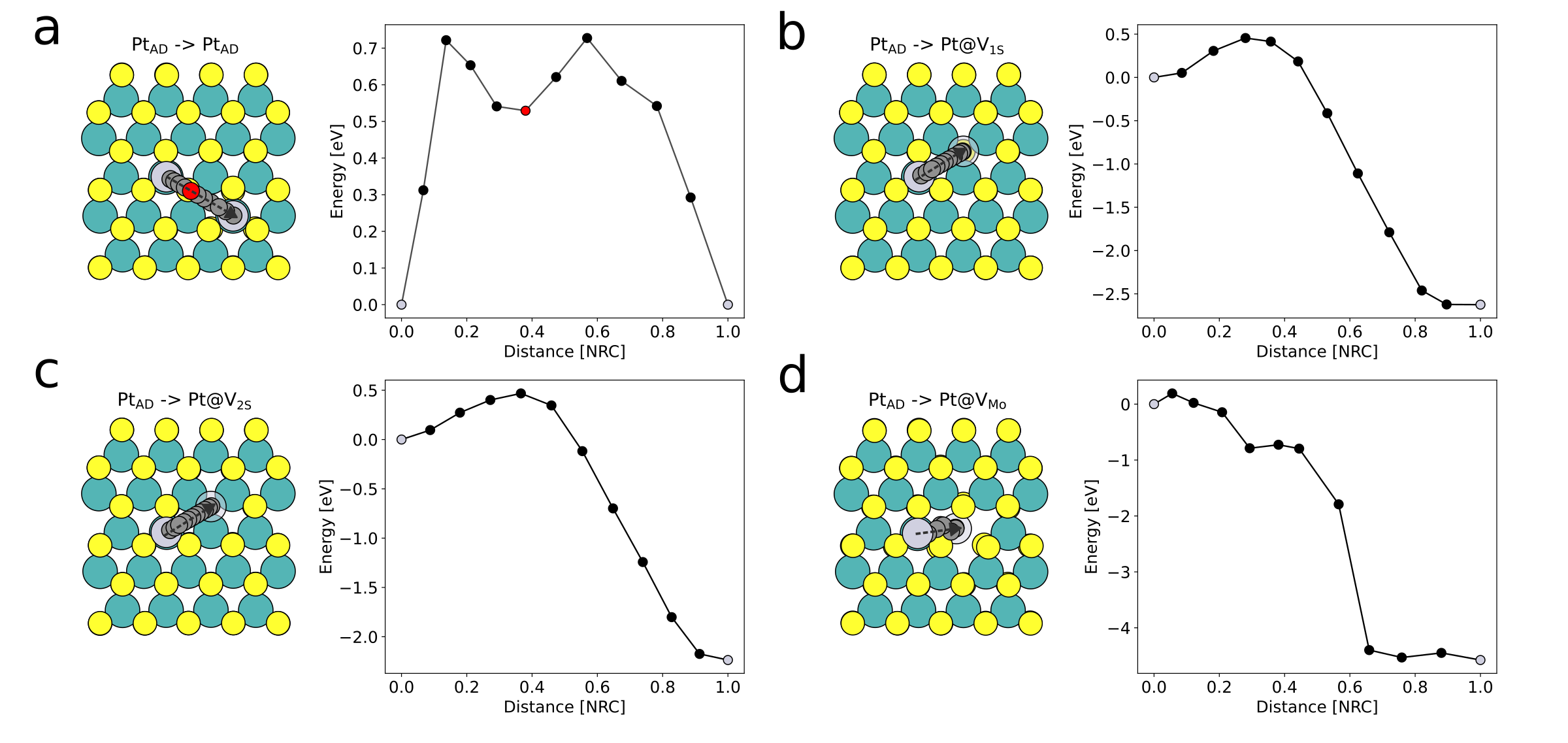}
    \caption{a-d) Diffusion paths of Pt atoms on the surface calculated by the nudged elastic band method. The \textit{x}-axis in the energy diagrams are given in relative atomic mass-weighted distances (normalized reaction coordinates NRC). The black dots in the energy diagram match the (semi-transparent) gray circles on the atomic model. Start and end positions of the diffusion process are marked with silver dots. a) Diffusion from the top of a Mo site to the top of another Mo site over the metastable position on top of a S site (marked by the red dot). b) Diffusion from the surface to a $\mathrm{V_{1S}}$ site, c) to a $\mathrm{V_{2S}}$ site and d) to a $\mathrm{V_{Mo}}$ site.}
    \label{Fig 3}
\end{figure}

As was already visible in Fig.~\ref{Fig 1}f, after Pt evaporation, most Pt atoms occupy the sites of the sulfur sublattice. Some of these Pt atoms are unstable under the electron beam and jump to another sulfur site once the scan reaches their position (see Supplementary Material Fig.~S6). These are most probably not adatoms on the pristine surface, as the energetically more favorable site for a Pt adatom is on top of a Mo site (see Fig.~\ref{Fig 3}a). Further, the surface diffusion migration barriers for Pt are only between 0.4 and 0.6 eV (see Fig.~\ref{Fig 3}a-b) depending on the location of the adatom, and thus the Pt atoms can easily diffuse over the \ce{MoS_2} surface until they fall into an energetically more favorable vacancy site. Therefore, we assume that the observed atom jumps take place between S vacancy sites, as described in Ref.~\citenum{li_atomic_2017}. The diffusion energy paths of Pt atoms into $\mathrm{V_{1S}}$, $\mathrm{V_{2S}}$ and $\mathrm{V_{Mo}}$ vacancies are depicted in Fig.~\ref{Fig 3}b-d and show binding energies of 2.6, 2.5 and 4.6~eV, respectively. All these binding energies are sufficient to ensure the stability of the implanted atom at room temperature.

\begin{figure}[htp]
    \centering
    \includegraphics[width= 14 cm]{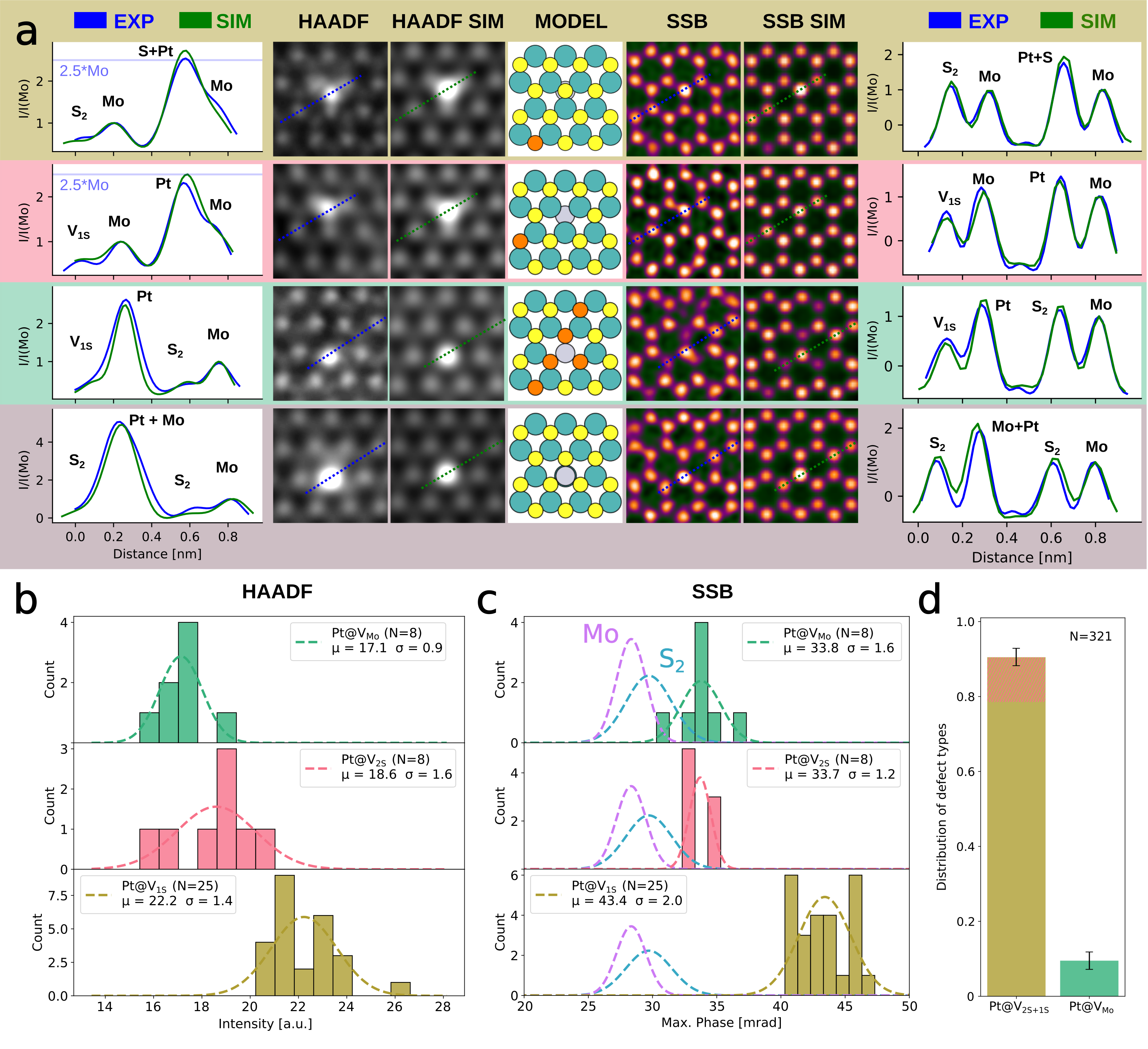}
    \caption{a) Line profiles of Gaussian blurred HAADF-STEM images of dopant structures, shown alongside the blurred images as well as HAADF image simulations, atomic models of the imaged structures, SSB reconstructions of the phase information at the same location, and realistic simulations of the SSB images. From top to bottom, the rows contain data of Pt@V$\mathrm{_{1S}}$, Pt@V$\mathrm{_{2S}}$, Pt@V$\mathrm{_{Mo}}$ and Pt adatom columns. b) Histograms of HAADF intensities at Pt@V$\mathrm{_{Mo}}$, Pt@V$\mathrm{_{2S}}$ and Pt@V$\mathrm{_{1S}}$ sites and Gaussian fits of the distributions. The types of the Pt substitutions at the  S$\mathrm{_2}$ sublattice are determined using the SSB phase at the respective locations. c) Histograms of SSB phase values at the same atomic sites as in b), together with Gaussian fits of the phase distribution of the structures. The violet and turquoise lines represent the distribution of the Mo and  S$\mathrm{_2}$ phase values, respectively. The $\mu$ in the legends indicates the center of the Gaussian fits, the $\sigma$ is the respective standard deviation. d) Relative occurrence of dopant structures obtained from large scale HAADF imaging. As Pt@V$\mathrm{_{1S}}$ and Pt@V$\mathrm{_{2S}}$ are barely distinguishable in HAADF images both types are counted together and the tip of the Pt$\mathrm{@V_{1S+2S}}$ column is shaded in red to mark the approximate ratio of Pt@V$\mathrm{_{2S}}$ based on the ratio between $\mathrm{V_{1S}}$ and $\mathrm{V_{2S}}$. The uncertainty in the columns is based on the variation between observed images and rounded up to the next integer percentage. \textit{N} is the number of cases for each histogram.}
    \label{Fig 4}
\end{figure}

Experimental and simulated HAADF and SSB images of the most typical Pt-doped sites are shown in Fig. \ref{Fig 4}a. As Pt atoms have a significantly higher nuclear charge (\textit{Z} = 78) than the surrounding Mo (\textit{Z} = 42) and S (\textit{Z} = 16) atoms, the Pt atom in HAADF images appears as a large bright feature, obscuring the neighboring atomic structure. The contrast difference between a Pt atom located at a $\mathrm{V_{2S}}$ site (Pt$\mathrm{@V_{2S}}$) with a theoretical Pt/Mo intensity ratio of 2.5 is very similar to Pt@V$\mathrm{_{1S}}$ with a theoretical (S+Pt)/Mo intensity ratio of 2.7. The SSB images give a much clearer picture, with a Pt/Mo phase ratio obtained by simulations of 1.19 and a respective simulated (S+Pt)/Mo phase ratio of roughly 1.85. This is reflected in the experimental data, where we observed an average phase of 43.4 $\pm$ $1.6$~mrad for Pt@V$\mathrm{_{1S}}$, 33.7 $\pm$ 1.1~mrad for Pt@V$\mathrm{_{2S}}$ and 33.8 $\pm$ 1.9~mrad for Pt@V$\mathrm{_{Mo}}$, which results in Pt/Mo and (S+Pt)/Mo ratios of 1.2 and 1.6, respectively. SSB imaging is also a powerful tool for analyzing Mo substitutions. Due to the considerably lower phase ratio of the Pt atom in comparison to the neighboring sites, discerning the neighborhood and exact placement of the Pt atom implanted in the defect clusters around $\mathrm{V_{Mo}}$ sites becomes significantly more straightforward, as can be seen in the third row of Fig.~\ref{Fig 4}a. 

Unfortunately the complex contrast formation in SSB may lead to misinterpretation of Pt adatoms located on top of the Mo sublattice, as can be seen in the fourth row of Fig.~\ref{Fig 4}a, where HAADF imaging of this configuration provides significantly better contrast with a theoretical (Pt+Mo)/Mo intensity ratio of 5.6 compared to a Pt/Mo ratio of 2.5. As the observed adatoms are not displaced by the electron beam, it is safe to assume that these Pt atoms are stabilized by very thin carbon contamination on the \ce{MoS_2} surface.

Figures~\ref{Fig 4}b and~\ref{Fig 4}c show histograms of the maximum HAADF intensities and phase maxima at the same Pt-doped sites. While Pt@V$\mathrm{_{1S}}$ and Pt@V$\mathrm{_{2S}}$ distributions overlap in HAADF imaging, the phase distributions in SSB imaging are significantly more distinct. As we have only obtained a limited amount of SSB data, the distribution of different Pt dopant types plotted in Fig.~\ref{Fig 4}d is based on large-scale HAADF images. Due to the poor distinguishability of Pt@V$\mathrm{_{1S}}$ and Pt@V$\mathrm{_{2S}}$ in HAADF, these are shown in the same column. Notably, the ratio between Pt atoms in S and Mo vacancies is the same as the ratio of the vacancies themselves, leading to the assumption that the Pt atoms are incorporated into the first defect they find after landing on the \ce{MoS_2} surface. Therefore we shaded the tip of the Pt@$\mathrm{V_{1S+2S}}$ column in Fig.~\ref{Fig 4}d in red to mark the approximate ratio of Pt@V$\mathrm{_{2S}}$ based on the ratio between $\mathrm{V_{1S}}$ and $\mathrm{V_{2S}}$.

Since the phase contrast is directly related to the local electron charge density, which can change depending on the chemical interactions between the atoms, using a model of non-interacting atoms is, strictly speaking, not sufficient for fully quantitative image simulations~\cite{madsen_ab_2021, hofer_detecting_nodate}. Supplemental Material Fig.~S7 contains simulated images with independent-atom-model (IAM) and DFT potentials of Pt@V$\mathrm{_{1S}}$ and Pt@V$\mathrm{_{2S}}$ structures as well as their difference. Evidently the simulated phase shifts due to charge transfer to the Pt dopant sites in \ce{MoS_2} are in the range of one percent of the absolute phase values and thus well within the range of the observed phase uncertainty in SSB. This suggests that simulations based on the IAM are precise enough for qualitative analysis in this system.

\section*{Conclusions and Outlook}
In summary, by combining helium ion irradiation to create controlled vacancy defects and subsequent Pt atom incorporation via evaporation, we successfully achieved substitutional platinum doping of sulfur ($\mathrm{V_{1S}}$, $\mathrm{V_{2S}}$) and molybdenum ($\mathrm{V_{Mo}}$) lattice sites in monolayer \ce{MoS_2}. We further demonstrated that SSB is a powerful imaging technique for reliably identifying and characterizing defect and dopant structures in Pt-doped \ce{MoS_2} monolayers at atomic resolution. The phase contrast obtained by SSB allows to reliably differentiate between various dopant and defect configurations, such as Pt atoms in single and double sulfur vacancies, which are difficult to resolve using HAADF-STEM imaging alone.

However, SSB is not without limitations: its phase contrast depends on the local atomic environment, there is a reduced $Z$-contrast between heavy and light elements, and further non-linearity can be introduced due to charge redistribution, all of which can make image interpretation challenging. It is therefore vital that SSB studies of modified 2D materials should be assisted by quantitative image simulations. It should also be noted that in some systems, SSB imaging does not provide any advantage over traditional imaging techniques, e.g., MoS$_2$ co-doped with two similar high $Z$-elements such as Au and Pt. Further, the larger data volumes and computational demands required for SSB make it less scalable for large-area imaging compared to HAADF-STEM. 

In the future, our substitutional doping method could be further optimized by fine-tuning the ion beam parameters to increase the precision of the defect creation process. Moreover, a resulfurization step could be applied before or after Pt evaporation to repair undesired defects, thereby improving the control and uniformity in dopant placement. These refinements enhance the scalability and reproducibility of Pt doping in \ce{MoS_2}, offering a promising pathway for controlled defect engineering and functionalization of materials for advanced applications in catalysis and electronics. Future experiments with substitutional dopants in MoS$_2$ could benefit from advanced 3D structural characterization techniques, such as few-tilt ptychotomography~\cite{hofer_picometer-precision_2023}, to further expand the potential of this approach for targeted material design.

\begin{suppinfo}

The following files are available free of charge:
\begin{itemize}
  \item Supplementary Material: Additional information about the sample preparation and the substitutional doping method, details of the experimental imaging conditions; remarks on contamination and EELS spectra; details of the image simulation as well as information about DFT calculations; supporting figures of the implantation process as well as an example diffraction pattern. 
  
\end{itemize}

\end{suppinfo}

\begin{acknowledgement}

We thank Neill McEvoy for kindly providing us with the~\ce{MoS_2} samples used in this study as well as Yossarian Liebsch and Umair Javed for providing \ce{MoS_2} samples that were used in preparation for this study. We acknowledge funding from the Austrian Science Fund (FWF) [10.55776/P35318, 10.55776/DOC142, 10.55776/36264, 10.55776/COE5].

\end{acknowledgement}

\subsection*{Author contributions}
DL, JK and LF conceived the study. DL and ML conducted the experimental work, TS provided code for SSB and DL, ML, TS, JK and LF performed the data analysis. AB and MC performed the \textit{ab initio} calculations. DL, JK, ML and LF wrote the manuscript, with contributions from all other authors. LF and JK supervised the study.

\subsection{Data and Code Availability Statement}
The experimental data supporting these findings will be made available through University of Vienna PHAIDRA repository upon acceptance of the manuscript. The SSB reconstruction code can be downloaded from \url{https://gitlab.com/pyptychostem/pyptychostem}

\subsection*{TOC graphic}

\begin{figure}[h!]
    \centering
    \includegraphics[]{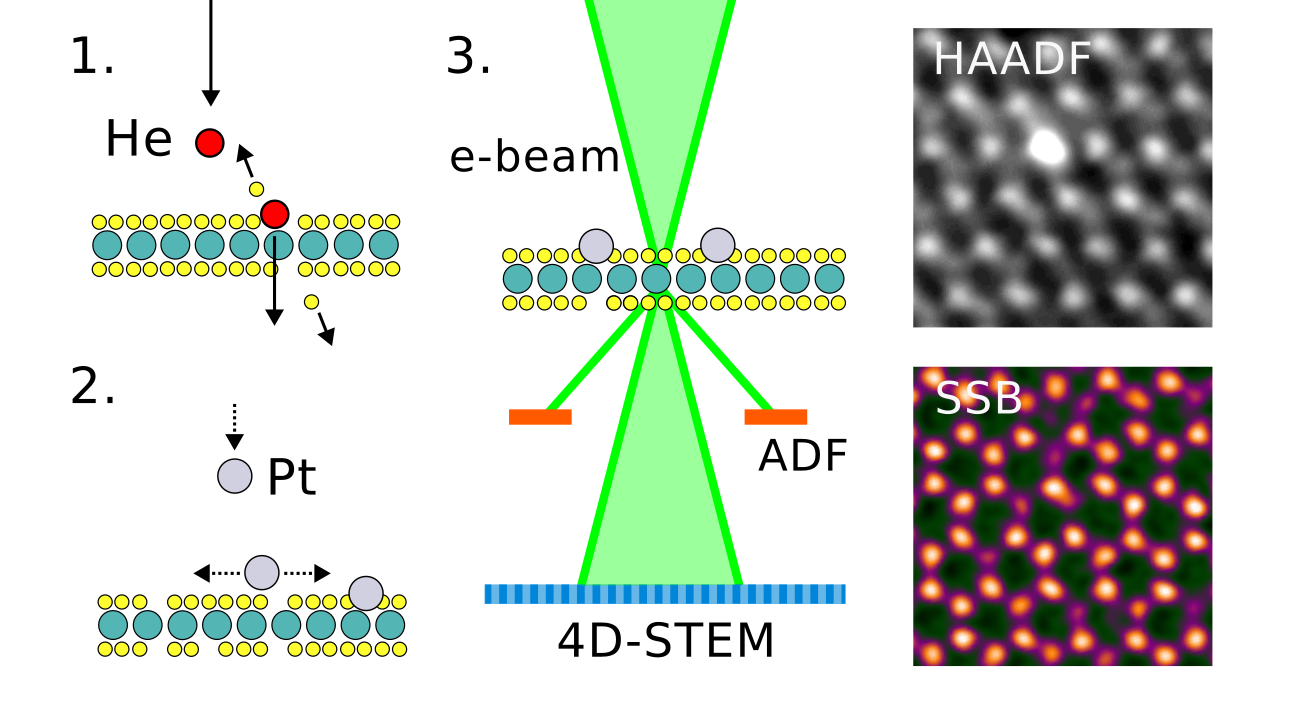}

\end{figure}

\bibliography{references}

\end{document}


\pagenumbering{gobble}
\pagenumbering{arabic}
\renewcommand{\appendixname}{Supplementary Material}
\renewcommand{\thefigure}{S\arabic{figure}} \setcounter{figure}{0}
\renewcommand{\thetable}{S\arabic{table}} \setcounter{table}{0}
\renewcommand{\theequation}{S\arabic{table}} \setcounter{equation}{0}

\section*{Methods}
\textbf{Sample preparation}
The \ce{MoS_2} sample was grown on \ce{SiO2} via chemical vapor deposition (CVD)~\cite{obrien_transition_2014} using liquid-phase exfoliated  \ce{MoO_3} as precursor and consists of mostly triangular monolayered flakes with an edge lengths between 5 and 20 $\mu$m. The samples were subsequently transferred in air onto a gold transmission electron microscopy grid with a holey membrane of amorphous carbon (Quantifoil R 1.2/1.3 Au grid) using the method described in Ref.~\citenum{meyer_hydrocarbon_2008}.

\textbf{Substitutional doping with single Pt atoms}
After transferring to the TEM grid, the MoS$_2$ samples were introduced into the interconnected near-UHV CANVAS system~\cite{mangler_materials_2022} (base pressure of 10$^{-8}$ mbar), which features both a SPECS ECR-HO microwave plasma generator and evaporation sources. Low-energy He ions from the plasma generator with a current of ca.~2.5~nA  at a He partial pressure of ca.~2.5$\times$10$^{-5}$~mbar were used to irradiate the \ce{MoS_2} samples. The measured ion energy for these parameters is approximately normally distributed with a mean of ca. 171 eV and a standard deviation of ca. 21 eV. Irradiation for 10 min corresponds to a fluence of ca. 1.25 × 10$^{13}$ cm$^{-2}$. The plasma treatment was followed by evaporation of the platinum. For Pt, the tip of a 99.9$\%$ Pt rod was heated up with an EFM-3 e-beam evaporator using a filament current of 2.9 A and an extraction voltage of 1850 V to produce a Pt flux of ca. 0.6 nA. After 10 min of evaporation, single Pt atoms were found in the \ce{MoS_2} lattice. 

\textbf{Microscopy and spectroscopy}
After implantation, the samples were transferred inside the UHV system to the aberration-corrected Nion UltraSTEM 100 scanning transmission electron microscope operated at 60 kV acceleration voltage with a beam current of ca. 40 pA. Images were acquired using a HAADF detector with a probe convergence semi-angle of ca. 35 mrad and a semi-angular range of 80–300 mrad. HAADF images with 2048$\times$2048 pixels were further processed using Gaussian blurring in order to reduce noise and increase contrast of the images. EEL spectra were recorded with a Gatan PEELS 666 spectrometer with an Andor iXon 897 CCD camera and an energy-dispersion of 0.5 eV/pixel~\cite{susi_single-atom_2017}. For simultaneous 4D data collection, stacks with 512$\times$512 real-space pixels were collected using a Dectris ARINA direct-electron detector (192 $\times$ 192 pixel) with a dwell time of 20 $\mathrm{\mu s}$ and an average dose of ca. $1 \times 10^{5}$ $\mathrm{e^-}$/\AA$^2$. The direct-electron detector falls completely within the HAADF detector and the maximum scattering angle collected by the detector is 36 mrad. To reduce the size of the data, recorded at each imaged position (which can be up to 10 GB for a single dataset), we reduced the size of the convergent-beam electron diffraction patterns recorded at each probe position to 48$\times$48 pixels by 4 times binning, which only has a negligent influence on the quality of the reconstructed phase~\cite{susi_open-source_2024}. After binning the reciprocal pixel size is approximately 1.5 mrad/pixel. SSB was performed with the open-source PyPtychoSTEM package~\cite{hofer_detecting_2023}, using the experimental 4D-STEM data as input. The convergence angle was set to 35 mrad and the step size was 0.156 \AA~per pixel (matched in the SSB reconstruction). Post-collection aberration correction was applied using singular value decomposition to identify the residual aberrations which were then counteracted.

\textbf{Density functional theory}
To determine the minimum energy paths and transition states for several possible diffusion events on the \ce{MoS_2} monolayer, a density functional theory (DFT) based \textit{ab initio} simulation approach as implemented in the CP2K~\cite{kuhne_cp2k_2020} code (version 2023.2) was used. 
For all calculations the norm-conserving, separable, dual-space Gaussian-type pseudopotentials of Goedecker, Teter, and Hutter (GTH)~\cite{goedecker_separable_1996} were used with a 500 Ry plane wave cutoff of the multigrid. The exchange-correlation interactions where treated with the Perdew–Burke–Ernzerhof (PBE)~\cite{perdew_generalized_1996} generalized gradient approximation. As a basis for all calculations, $ 5 \times 5 \times 1 $ \ce{MoS_2} monolayer supercells were used consisting of $75$ atoms with a vacuum layer of 40 \AA \, in order to prevent artificial interactions between periodic images. After a cell relaxation of the pristine material, the supercells were modified in order to determine the minimum energy paths for diffusion of Pt and S atoms on the surface of (defective) \ce{MoS_2}. For this, the climbing image nudged elastic band method~\cite{henkelman_climbing_2000} was used with 12 images and a maximum force convergence criteria of 0.01 Hartree per Bohr.

\textbf{HAADF and 4D-STEM image simulations}
HAADF-STEM image simulations were performed based on the DFT-relaxed models using the \textit{ab}TEM package~\cite{madsen_abtem_2021}. Similar to the experiment, the HAADF detector semi-angular ranges was set to 80–300 mrad and the probe convergence angle was set to 35 mrad. For all simulations an electron beam energy of 60 keV was assumed. To account for thermal diffuse scattering, we implemented the frozen-phonon model with 20 snapshots per image using standard deviation of atomic displacements values taken from Ref.~\citenum{mannebach_dynamic_2015}. For 4D-STEM the same parameters were used, but instead of the HAADF detector a pixelated detector setting was used. After creating the artificial 4D-STEM data, the phase images were reconstructed using the same algorithm and parameters as in the corresponding experimental images, including the dose per area simulated by adding Poisson noise to the diffraction patterns. To account for finite probe-size effects in SSB images, we added a Gaussian blur over the simulated images to match the line profile of the experimental SSB data. For the charge-transfer simulations, the DFT potential was calculated from the all-electron charge density converged with GPAW, as described in Ref.~\citenum{susi_efficient_2019}.

\textbf{Evaporation of Pt on graphene and EELS measurements}
Commercial graphene grown via CVD provided on a sacrificial polymethyl methacrylate (PMMA) layer (Easy Transfer graphene) was transferred onto a SiN TEM grid with with 3 $\mathrm{\mu m}$ holes (Silson Ltd) via a liquid transfer method using deionized water as the carrier liquid.
Following the  transfer onto the substrate the graphene was heated on a hot plate at 150$^{\circ}$C for 1~h. The PMMA sacrificial layer  was then removed via an acetone bath at 50$^{\circ}$C for 1~h, after which the grid rested in isopropyl alcohol (IPA) at room temperature for another hour.
Graphene samples were cleaned using a 6~W  continuous wave diode laser with a wavelength of 445~nm with a spot size of 0.3$\times$1.5~mm$^2$ as described in Ref.~\citenum{trentino_two-step_2022}. The surface was illuminated at 27\% of the maximum power for 6~min multiple times, where the power was set by changing the duty cycle. With these parameters, not all of the surface contamination is removed, but large-enough atomically clean patches are created for acquiring statistically meaningful data. Pt was then evaporated onto the sample for 15~minutes with a flux of 0.25~nA.

\section*{Remarks on contamination}

A significant number of the evaporated metal atoms can also be found on or next to hydrocarbon contamination (brighter, diffuse contrast on the image). Most of these atoms are still isolated and perhaps form chemical bonds with the hydrocarbons~\cite{li_atomic_2017}, but are not incorporated into the lattice. 

As is evident from the images in main article Fig. 1 d-f, ion irradiation and metal evaporation lead to a slight increase in undesired hydrocarbon contamination of the sample, effectively covering vacancy sites and incorporated metals. The same can be expected for  exposure of the defective \ce{MoS_2} surface to ambient conditions during sample transfer. To allow efficient implantation of dopants, protecting the sample from ambient between the steps is therefore crucial for the procedure. Nevertheless, even in UHV, every modification step can create additional contamination. Contamination during evaporation can be minimized by heat-treating the Pt source and cleaning the modification chamber with O$_2$ plasma (see Supplementary Material, Fig.~\ref{Fig Contamination 1}).  

\section*{EELS spectra of the evaporated material}

Eeven though the high \textit{Z}-contrast of the dopant atoms in HAADF-STEM and the excellent agreement with SSB image simulations are strong evidence for the element of the implanted atoms, definitive proof of the implantation of Pt atoms could benefit from additional spectroscopic evidence. Unfortunately, we were not able to obtain single-atom electron-energy loss (EEL) spectra of the Pt \textit{\ce{O_2}} edge (the only Pt edge accessible to our EEL spectrometer) due to the limited stability of the \ce{MoS_2} under the electron beam and the low cross section of the Pt edge. Nevertheless, to provide spectroscopic evidence of the evaporated material, we evaporated Pt atoms onto graphene using the same setup and similar parameters. Supplementary Material Fig.~\ref{Fig EELS} shows EEL spectra of a small Pt cluster formed in the graphene lattice, unambiguously demonstrating the presence of Pt.

\section*{Supplementary Figures}

\begin{figure}[htp]
    \centering
    \includegraphics[width= 16 cm]{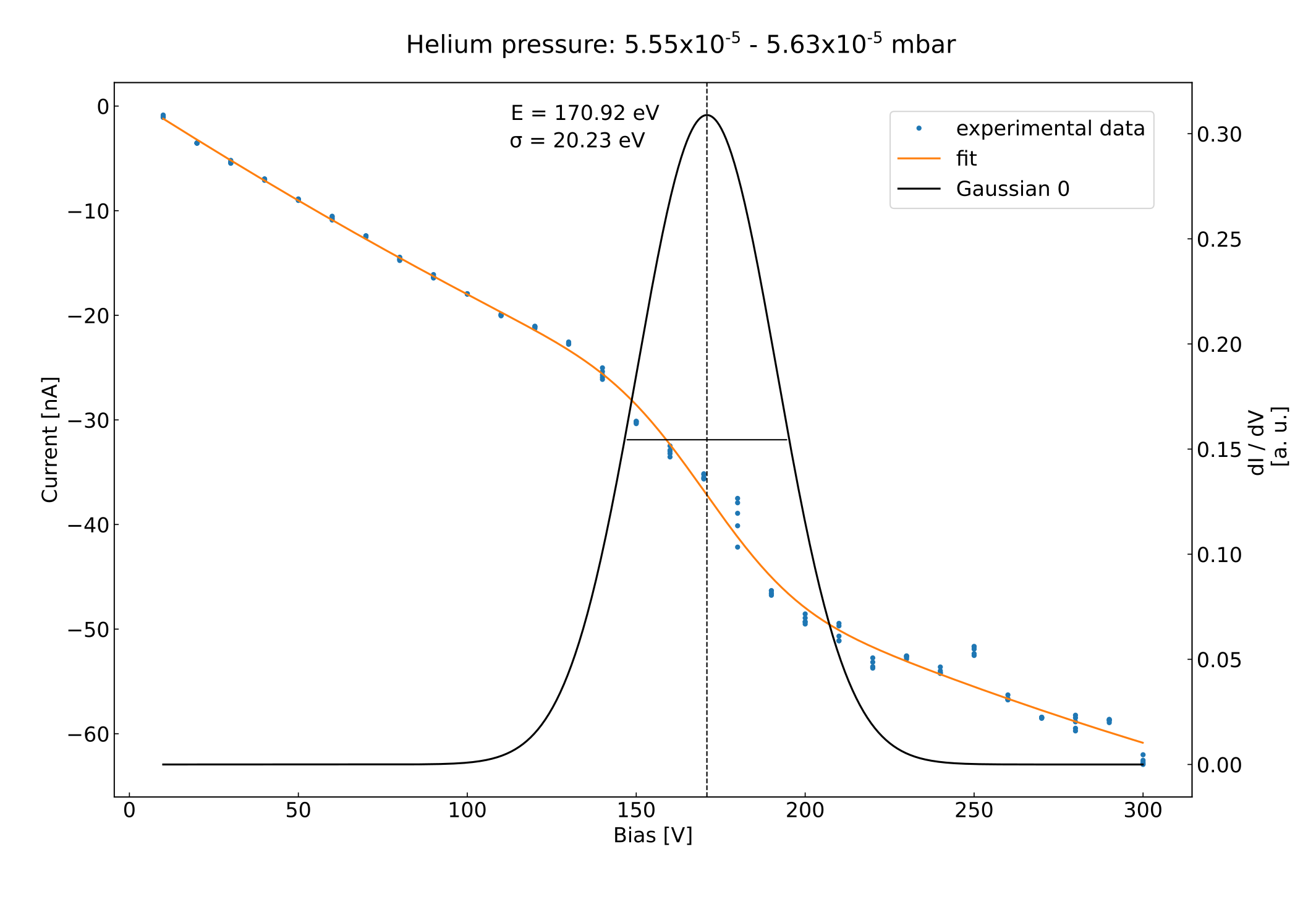}
    \caption{Beam profile and d\textit{I}/d\textit{V} curve of the deceleration measurement of the ion plasma. The experimental details of this analysis can be found in \cite{langle_defect-engineering_2024}.} 
    \label{Fig Beam Profile}
\end{figure}

\begin{figure}[htp]
    \centering
    \includegraphics[width= 16 cm]{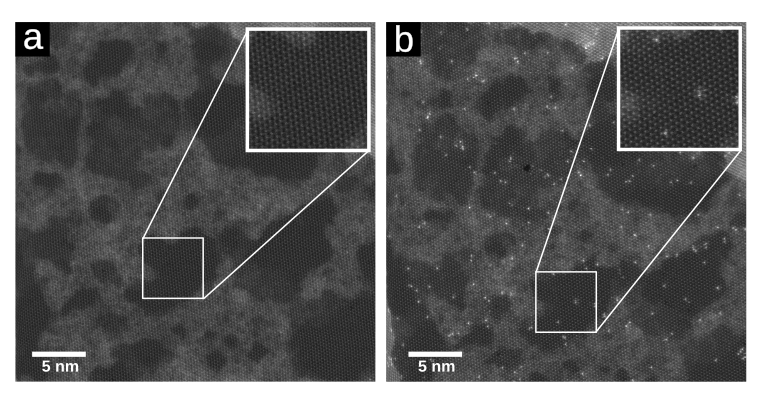}
    \caption{a) HAADF-STEM image of the pristine \ce{MoS_2} surface before evaporation b) HAADF-STEM image of the same area after evaporation of single Pt atoms. } 
    \label{Fig Contamination 1}
\end{figure}

\begin{figure}[htp]
    \centering
    \includegraphics[width= 12 cm]{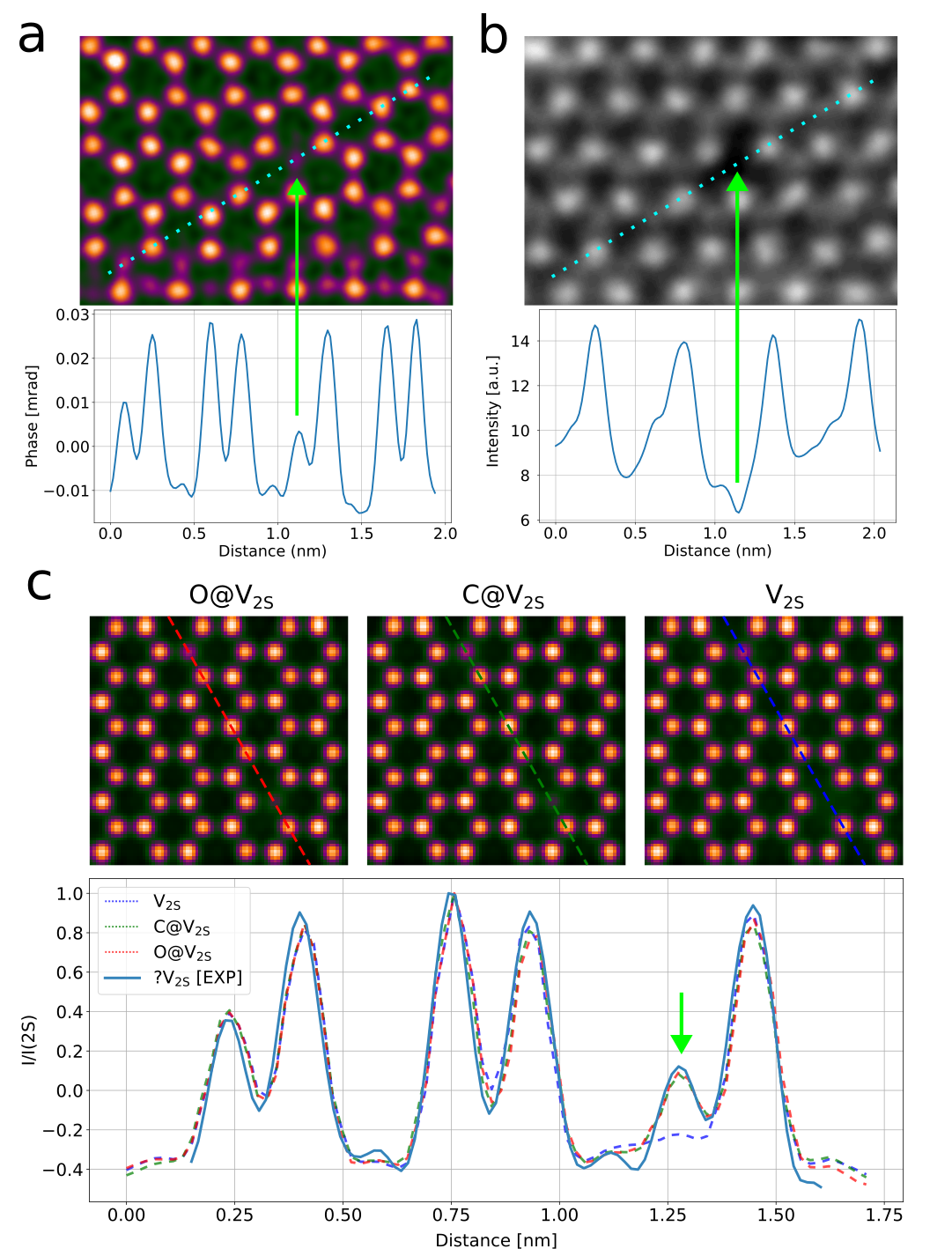}
    \caption{a) SSB phase image of highly defective \ce{MoS_2} next to a single vacancy defect line similar to the one reported in Ref.~\citenum{komsa_point_2013}. The line profile over a $\mathrm{V_{1S}}$ and a $\mathrm{V_{2S}}$ is horizontally matched with the image. The unexpected local maximum at the $\mathrm{V_{2S}}$ is marked with a green arrow. b) Gaussian blurred HAADF-STEM image of the same structure with horizontally matched line profile. The HAADF-STEM contrast has a minimum exactly at the location of the $\mathrm{V_{2S}}$ (green arrow). c) Simulated SSB phase images of $\mathrm{V_{2S}}$ without heteroatoms, with a C dopant, and a O dopant. The plot below contains line profiles of the simulated structures and the experimental data from panel a. The location of the questionable $\mathrm{V_{2S}}$ site is marked with a green arrow in the line profile.} 
    \label{Fig Wrong}
\end{figure}

\begin{figure}[htp]
    \centering
    \includegraphics[width= 11 cm]{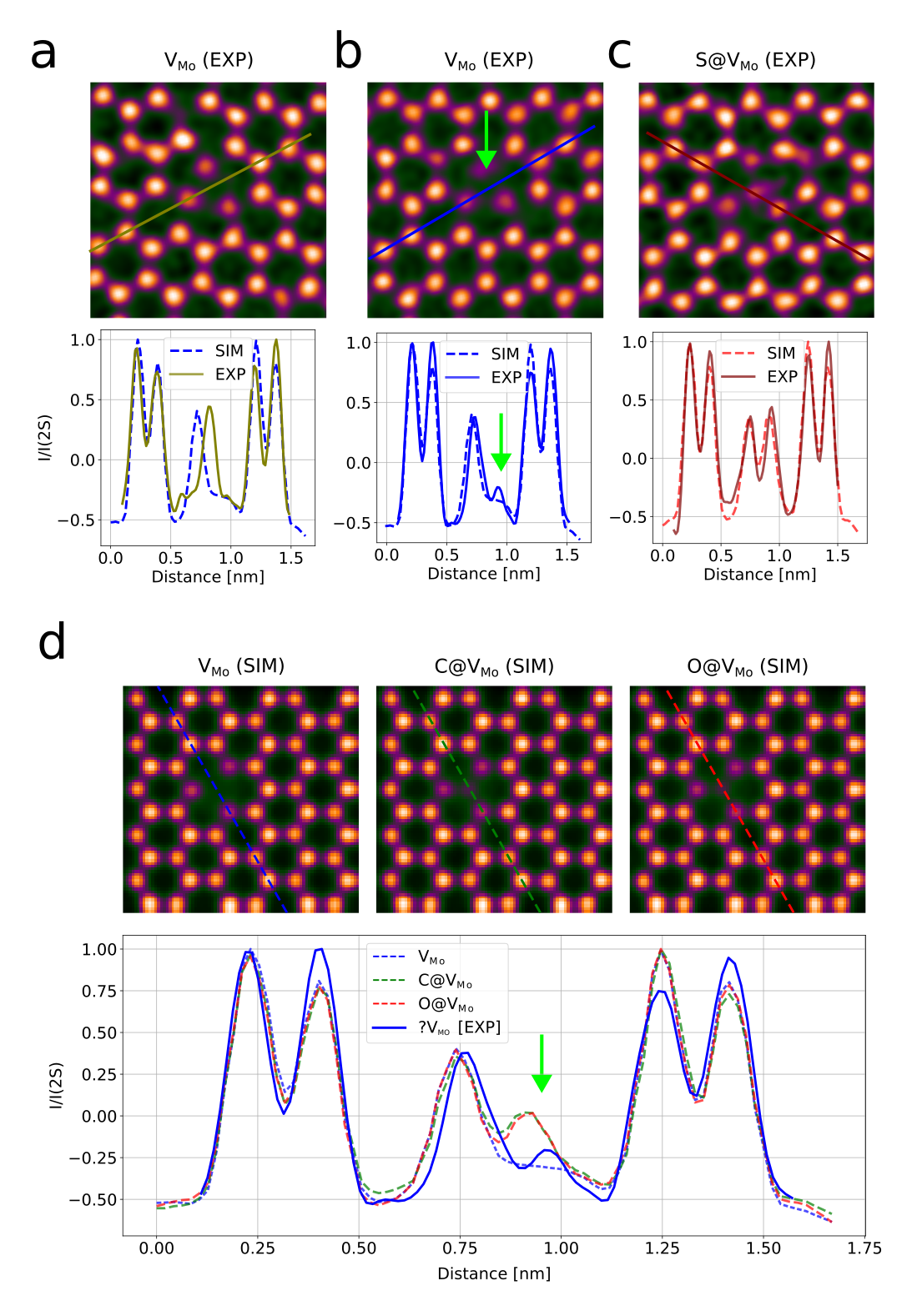}
    \caption{a) SSB phase image of a defect cluster with three S atoms missing around a $\mathrm{V_{Mo}}$ site. The simulated line profile corresponds to the first image in panel d of this figure. b) SSB phase of a similar defect cluster as in a. The simulated line profile corresponds to the first image in panel d. The unexpected local maximum at the $\mathrm{V_{Mo}}$ is marked with a green arrow. c) SSB image of a defect cluster with a S atom replacing the Mo atom. The simulated line profile corresponds to an S atom at the Mo position. d) Simulated SSB phase images of $\mathrm{V_{Mo}}$ surrounded by three $\mathrm{V_{1S}}$ without heteroatoms, with a C dopant and O dopant. The plot below contains line profiles of the simulated structures and the experimental data from panel b. The location of the questionable $\mathrm{V_{Mo}}$ site is marked with a green arrow in the line profile.} 
    \label{Fig Mo}
\end{figure}

\begin{figure}[htp]
    \centering
    \includegraphics[width= 16 cm]{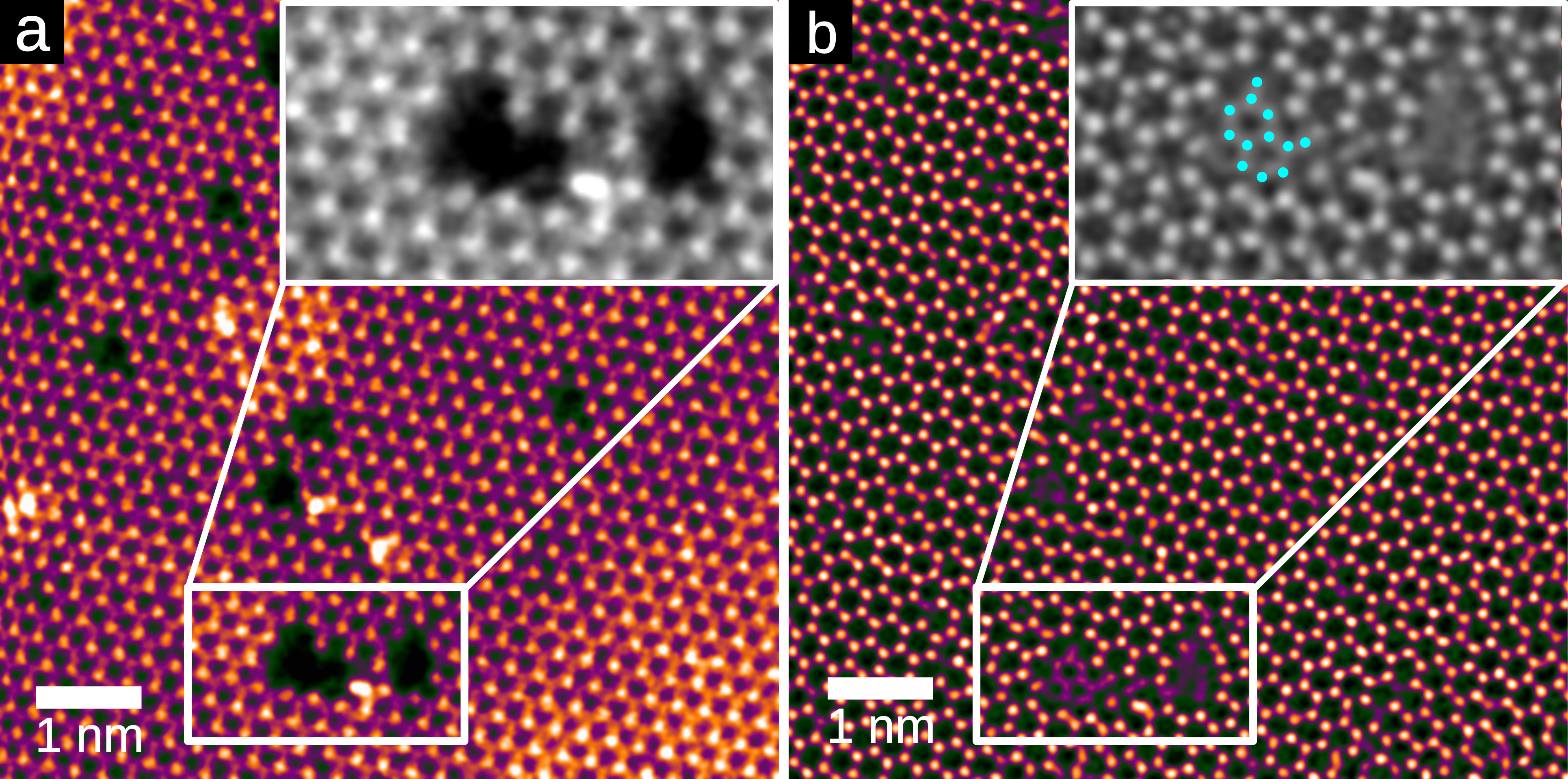}
    \caption{a) HAADF-STEM image of a defect cluster on contaminated \ce{MoS_2}. b) SSB phase image of the same area: inside the defect cluster a hexagonal ring of $sp²$ hybridized carbon is clearly visible. The approximate positions of the carbon atoms in the defect cluster are marked in the inset with cyan dots.} 
    \label{Fig Contamination}
\end{figure}

\begin{figure}[htp]
    \centering
    \includegraphics[width= 16 cm]{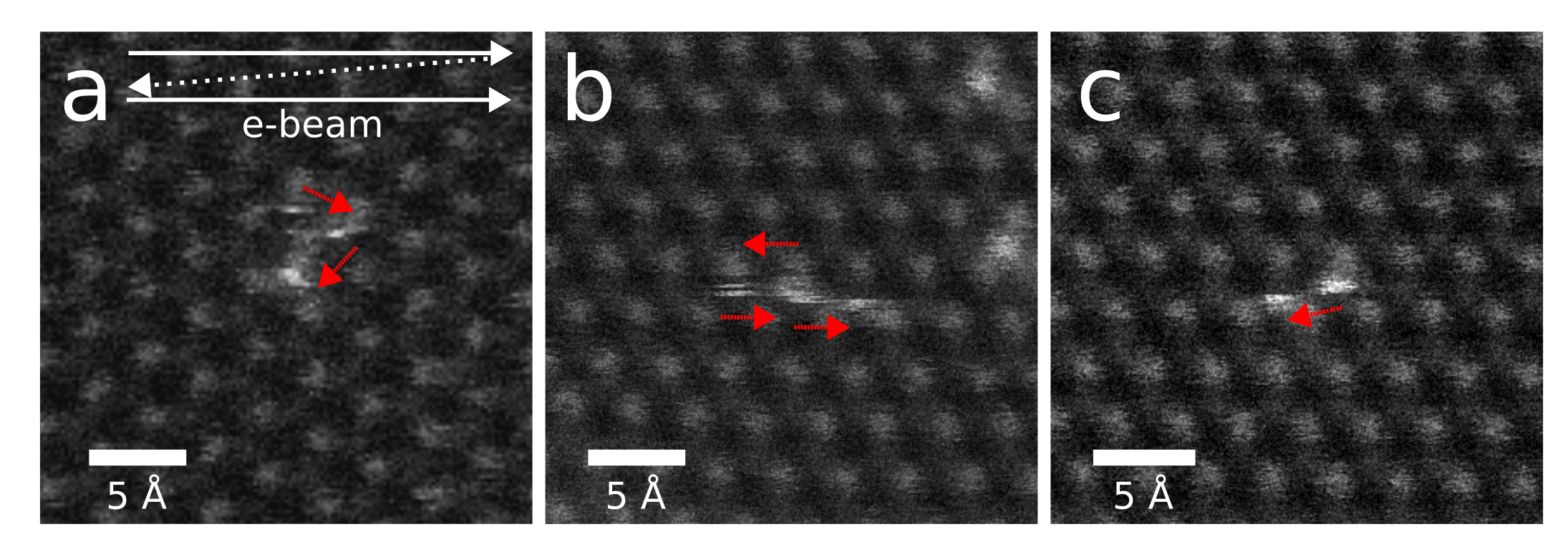}
    \caption{a-c) HAADF-STEM images of Pt atoms that are displaced from their position by the electron beam and imaged at different locations within the same frame. The trajectory of the Pt atoms is marked in red. The direction of the scanning electron beam (white lines) is overlaid in panel a).}
    \label{Fig Jumping}
\end{figure}

\begin{figure}[htp]
    \centering
    \includegraphics[width= 16 cm]{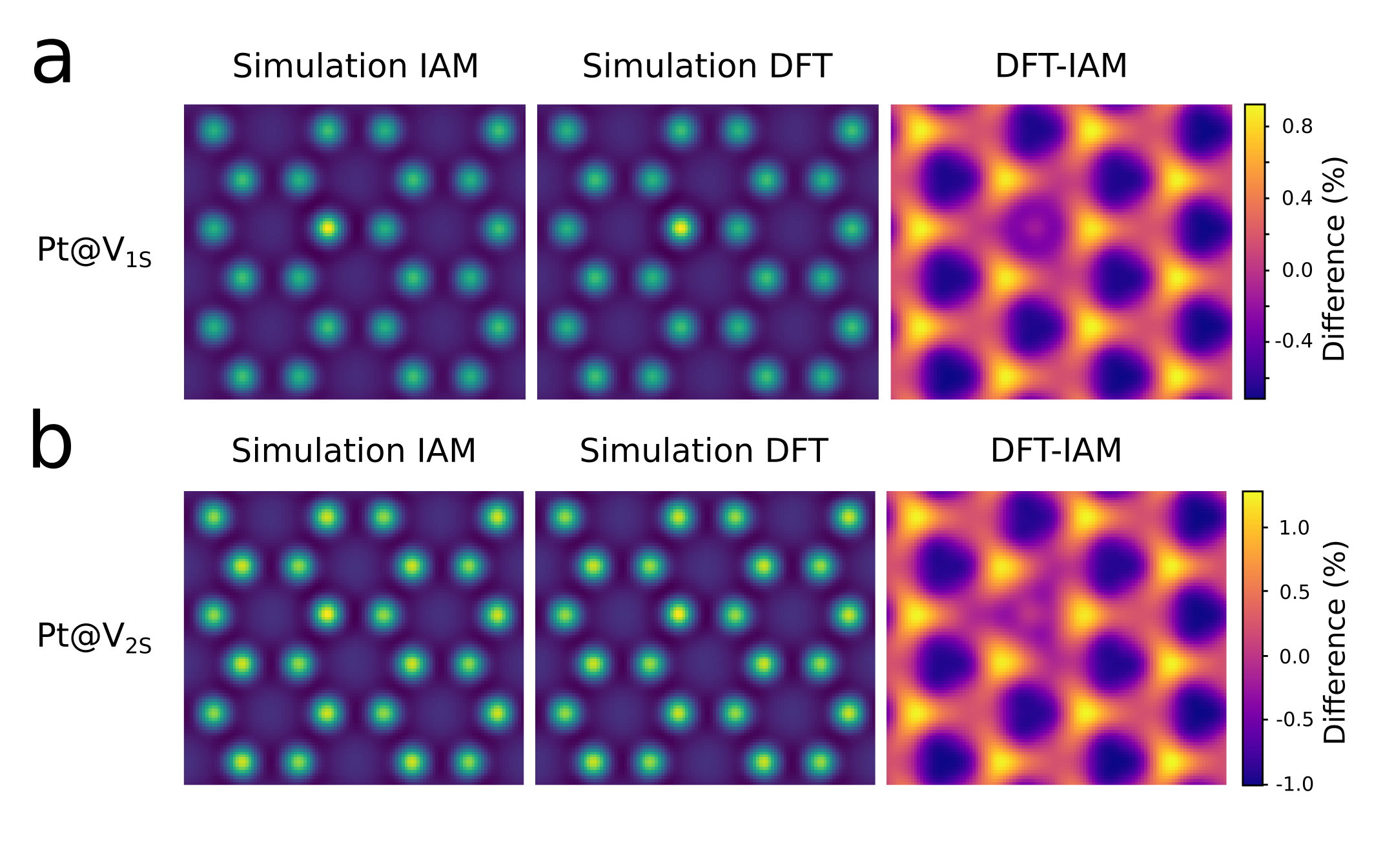}
    \caption{Phase image simulations of structures with potentials based on IAM and DFT as well as their difference for a) $\mathrm{Pt@V_{1S}}$ and b) $\mathrm{Pt@V_{2S}}$.} 
    \label{Fig CT}
\end{figure}

\begin{figure}[htp]
    \centering\includegraphics[width= 16 cm]{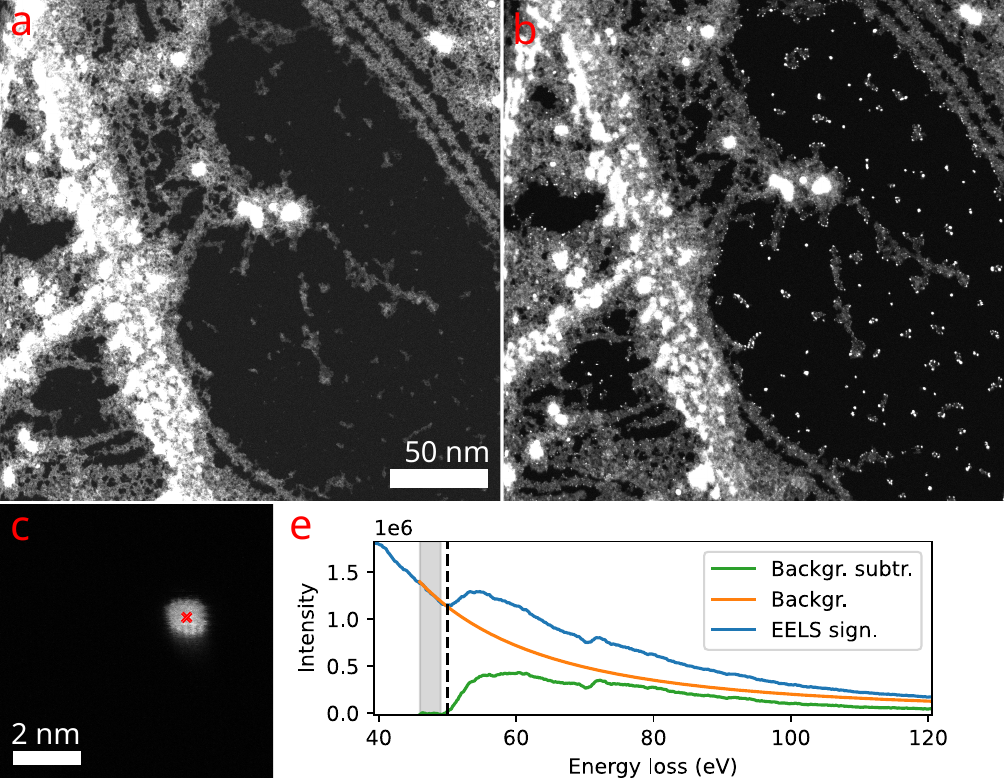}
    \caption{a) Overview HAADF-STEM image of laser-cleaned graphene.
             b) Overview HAADF-STEM image of the same area after Pt evaporation with an e-beam evaporator (flux: 0.25~nA, time: 15~min). 
             c) Smaller field of view HAADF-STEM image of Pt cluster on graphene with the position of the EEL spectrum (shown in d) marked with a red cross.
             d) EEL spectrum recorded at the position of the red cross in panel c. The signal of the Pt cluster is shown in blue, with the background in orange and the background-subtracted spectrum in green.
             A power law is fitted to the original spectrum in the range of 46.0 to 49.0~eV (gray area).
             A black marker is positioned at an energy-loss of 50 eV.
             This corresponds to the Pt~\textit{O}$_{2,3}$ and Pt~\textit{N}$_{4,6}$ edges.
             Spectral analysis was done using hyperSpy~\cite{pena_hyperspyhyperspy_2024}.
             The peaks align with reference spectra (EELS Atlas, https://eels.info/atlas/platinum).
             This also agrees closely with the spectrum shown in the Supplementary Figure S23 of Ref.~\citenum{campos-roldan_structure_2023}
            } 
    \label{Fig EELS}
\end{figure}

\begin{figure}[htp]
\centering\includegraphics[width= 16 cm]{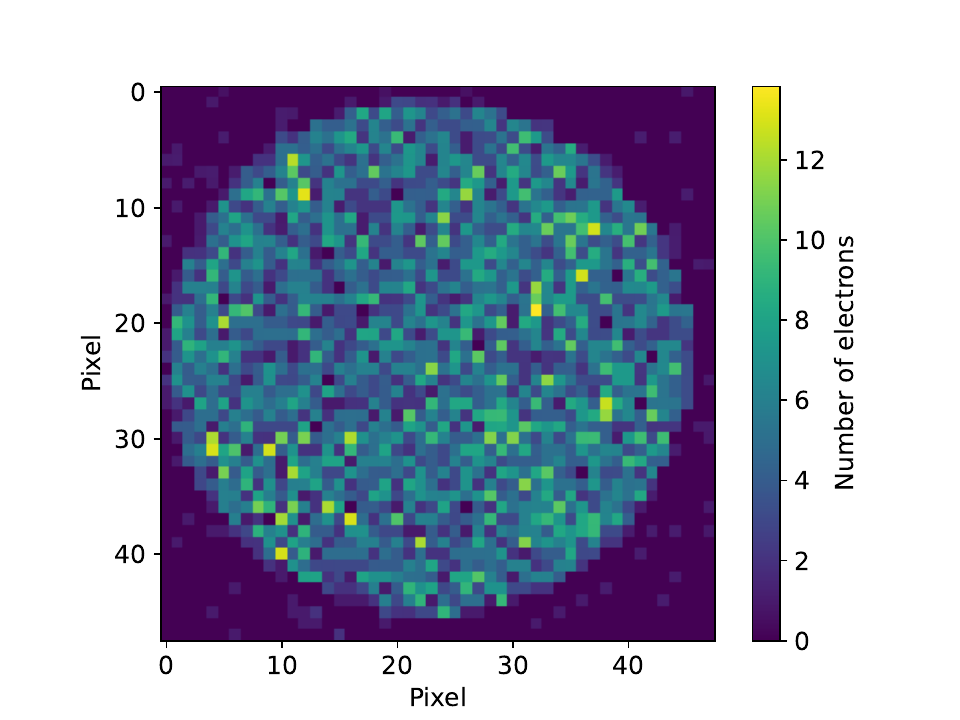}
\caption{Typical convergent-beam diffraction pattern obtained with the Dectris ARINA direct-electron detector after 4-times binning. } 
\label{Fig Diffraction}
\end{figure}
\clearpage
\bibliography{references}